\documentclass[11pt]{article}

\usepackage{amscd}
\usepackage{amsmath}

\newcommand{\N}{N\raise.7ex\hbox{\underline{$\circ $}}$\;$}

\begin{document}

\begin{center}
{\bf Tokarevskaya N.G., Ovsiyuk E.M., Red'kov V.M. \\[3mm]
MAXWELL EQUATIONS IN COMPLEX FORM, SPHERICAL WAVES \\ IN SPACES OF
CONSTANT CURVATURE OF LOBACHEVSKY AND RIEMANN }

\end{center}

\begin{quotation}

Complex formalism of Riemann -- Silberstein -- Majorana --
Oppenheimer in Maxwell electrodynamics is extended to the case of
arbitrary pseudo-Riemannian space -- time in accordance with the
tetrad recipe of Tetrode -- Weyl -- Fock -- Ivanenko. In this
approach, the Maxwell equations are solved exactly
 on the background of simplest static cosmological models, spaces of constant curvature of Riemann and Lobachevsky
  parameterized by spherical coordinates.
Separation of variables is realized in the basis of
Schr\"{o}dinger -- Pauli type, description of angular dependence
in electromagnetic complex 3-vectors is given in terms of Wigner
$D$-functions. In the case of compact Riemann model a discrete
frequency spectrum
 for electromagnetic modes depending on the curvature radius of space and three discrete
 parameters is found. In the case of hyperbolic Lobachevsky model
 no discrete spectrum for frequencies of electromagnetic modes arises.

\end{quotation}

\section{ Introduction: matrix complex form of Maxwell theory
}

The task of the present paper is to obtain in explicit form
spherical waves solutions to Maxwell equations in space of
positive and negative curvature, spherical Riemann $S_{3}$
Lobachevsky $H_{3}$ models, when they are parameterized by
extended spherical coordinates. This paper continues investigation
of similar problems on searching solutions of the Maxwell
equations in symmetrical space-time model \cite{2006-Bogush,
2006-Bychkovska}.

We will use the known complex form of Maxwell theory according to
approach by Riemann, Silberstein, Oppenheimer,
 and Majorana \cite{1901-Weber},
 \cite{1907-Silberstein(1)},\cite{1931-Oppenheimer},
\cite{1931-Majorana}, (also see [8--32]), which is extended to the
case of arbitrary curved space -- time in accordance with general
tetrad formalism by Tetrode -- Weyl -- Fock -- Ivanenko
 [33-35]; also see \cite{Book}).

\hspace{5mm} Let us start with Maxwell equations in vacuum at
presence of sources: (with the use of notation $ j^{a} = (\rho,
{\bf J} /c) \; , \; c^{2} = 1 / \epsilon_{0} \mu_{0} \; $):
\begin{eqnarray}
 \mbox{div} \; c{\bf B} = 0 \; , \qquad \mbox{rot}
\;{\bf E} = -{\partial c {\bf B} \over \partial ct} \; , \nonumber
\\
 \mbox{div}\; {\bf E} = {\rho \over \epsilon_{0}} , \qquad
 \mbox{rot} \; c{\bf B} = {{\bf j} \over \epsilon_{0}} +
   {\partial {\bf E} \over \partial ct} \; .
\label{1.2a}
\end{eqnarray}

\noindent In explicit form they are
\begin{eqnarray}
\partial_{1} cB^{1} + \partial_{2} cB^{2} + \partial_{3} cB^{3} = 0 \; ,
\qquad
\partial_{2} E^{3} - \partial_{3} E^{2} +\partial_{0} cB^{1} =0 \; ,
\nonumber
\\
\partial_{3} E^{1} - \partial_{1} E^{3} + \partial_{0} cB^{2} = 0 \; ,
\qquad
\partial_{1} E^{2} - \partial_{2} E^{1} + \partial_{0} cB^{3} = 0\; ,
\nonumber
\\
\partial_{1} E^{1} + \partial_{2} E^{2} + \partial_{3} E^{3} = j^{0}/\epsilon_{0} \; ,
\qquad
\partial_{2} cB^{3} - \partial_{3} cB^{2} -\partial_{0} E^{1} =j^{1}/\epsilon_{0} \; ,
\nonumber
\\
\partial_{3} cB^{1} - \partial_{1} cB^{3} - \partial_{0} E^{2} = j^{2}/\epsilon_{0} \; ,
\qquad
\partial_{1} cB^{2} - \partial_{2} cB^{1} - \partial_{0} E^{3} = j^{3}/\epsilon_{0}\; .
\label{1.2b}
\end{eqnarray}

\noindent
 Let us introduce a complex 3-component vector $ \psi^{k}
= E^{k} + i c B^{k}$; eqs. (\ref{1.2b}) are easily combined into
 (for shortness $c=1$):
\begin{eqnarray}
\partial_{1}\Psi ^{1} + \partial_{2}\Psi ^{0} + \partial_{3}\Psi ^{3} =
j^{0} / \epsilon_{0} \; , \nonumber
\\
-i\partial_{0} \psi^{1} + (\partial_{2}\psi^{3} -
\partial_{3}\psi^{2}) = i\; j^{1} / \epsilon_{0} \; ,
\nonumber
\\
-i\partial_{0} \psi^{2} + (\partial_{3}\psi^{1} -
\partial_{1}\psi^{3}) = i\; j^{2} / \epsilon_{0} \; ,
\nonumber
\\
-i\partial_{0} \psi^{3} + (\partial_{1}\psi^{2} -
\partial_{2}\psi^{1}) = i\; j^{3} / \epsilon_{0} \; ,
\label{1.3}
\end{eqnarray}

\noindent or in  matrix form: (12 arbitrary parameters enter this
matrix equation):
\begin{eqnarray}
(-i\alpha^{0} \partial_{0} + \alpha^{j} \partial_{j} ) \Psi = J \;
, \qquad \Psi = \left | \begin{array}{c} 0 \\\psi^{1} \\\psi^{2}
\\ \psi^{3}
\end{array} \right | \; , \qquad
\alpha^{0} = \left | \begin{array}{rrrr}
a_{0} & 0 & 0 & 0 \\
a_{1} & 1 & 0 & 0 \\
a_{2} & 0 & 1 & 0 \\
a_{3} & 0 & 0 & 1
\end{array} \right | \; ,
\nonumber
\\
\alpha^{1} = \left | \begin{array}{rrrr}
b_{0} & 1 & 0 & 0 \\
b_{1} & 0 & 0 & 0 \\
b_{2} & 0 & 0 & -1 \\
b_{3} & 0 & 1 & 0
\end{array} \right | ,\;\;
\alpha^{2} = \left | \begin{array}{rrrr}
c_{0} & 0 & 1 & 0 \\
c_{1} & 0 & 0 & 1 \\
c_{2} & 0 & 0 & 0 \\
c_{3} & -1 & 0 & 0
\end{array} \right | ,\;\;
\alpha^{3} = \left | \begin{array}{rrrr}
d_{0} & 0 & 0 & 1 \\
d_{1} & 0 & -1 & 0 \\
d_{2} & 1 & 0 & 0 \\
d_{3} & 0 & 0 & 0
\end{array} \right | \; .
\label{1.4}
\end{eqnarray}

\noindent Taking into account
\begin{eqnarray}
(\alpha^{0} )^{2}=  \left | \begin{array}{rrrr}
a_{0}a_{0} & 0 & 0 & 0 \\
a_{1}a_{0} +a_{1} & 1 & 0 & 0 \\
a_{2}a_{0} +a_{2} & 0 & 1 & 0 \\
a_{3}a_{0} +a_{3} & 0 & 0 & 1
\end{array} \right | \; .
\nonumber
\end{eqnarray}

\noindent let us require
\begin{eqnarray}
(\alpha^{0} )^{2}= +I , \qquad a_{0}a_{0}=1\; , \;\; a_{1}a_{0}
+a_{1} \;, \;\; a_{2}a_{0} +a_{2}\; , \;\; a_{3}a_{0} +a_{3} \; ;
\nonumber
\end{eqnarray}

\noindent the most simple solution is
\begin{eqnarray}
a_{0}= \pm 1, \qquad a_{j}= 0 \; , \qquad \alpha^{0} = \left |
\begin{array}{rrrr}
\pm 1 & 0 & 0 & 0 \\
0 & 1 & 0 & 0 \\
0 & 0 & 1 & 0 \\
0 & 0 & 0 & 1
\end{array} \right | , \qquad (\alpha^{0} )^{2}= +I\; .
\label{1.5}
\end{eqnarray}

\noindent In similar manner, taking
\begin{eqnarray}
(\alpha^{1})^{2} =  \left | \begin{array}{rrrr}
b_{0}^{2}+b_{1} & b_{0} & 0 & 0 \\
b_{1}b_{0} & b_{1} & 0 & 0 \\
b_{2}b_{0}-b_{3} & b_{2} & -1 & 0 \\
b_{3}b_{0}-b_{2} & b_{3} & 0 & -1
\end{array} \right | \; .
\nonumber
\end{eqnarray}

\noindent for equation $ (\alpha^{1})^{2} = -I $ let us use the
most simple solution
\begin{eqnarray}
b_{0}=0 \; , \qquad b_{1}= -1 \; , \qquad b_{2}=0 \; , \qquad
b_{3}=0 \; , \qquad \alpha^{1} = \left | \begin{array}{rrrr}
0 & 1 & 0 & 0 \\
-1 & 0 & 0 & 0 \\
0 & 0 & 0 & -1 \\
0 & 0 & 1 & 0
\end{array} \right | \; .
\label{1.6}
\end{eqnarray}

\noindent Again,
\begin{eqnarray}
(\alpha^{2})^{2} = \left | \begin{array}{rrrr}
c_{0}c_{0}+c_{2} & 0 & c_{0} & 0 \\
c_{1}c_{0}+c_{3} & -1 & c_{1} & 0 \\
c_{2}c_{0} & 0 & c_{2} & 0 \\
c_{3}c_{0}-c_{1} & 0 & c_{3} & -1
\end{array} \right | = -I\; ,
\nonumber
\end{eqnarray}

\noindent that is
\begin{eqnarray}
c_{0}=0\; , \; c_{1}= 0 \; , \; c_{2}=-1 \; , \; c_{3}=0 \;,
\qquad \alpha^{2} = \left | \begin{array}{rrrr}
0 & 0 & 1 & 0 \\
0 & 0 & 0 & 1 \\
-1 & 0 & 0 & 0 \\
0 & -1 & 0 & 0
\end{array} \right | \; , \qquad (\alpha^{2})^{2} = -I\; .
\label{1.7}
\end{eqnarray}

\noindent And finally, for $\alpha^{3}$ we have
\begin{eqnarray}
(\alpha^{3})^{2} = \left | \begin{array}{rrrr}
d_{0}d_{0}+d_{3} & 0 & 0 & d_{0} \\
d_{1}d_{0}-d_{2} & -1 & 0 & 0 \\
d_{2}d_{0}+d_{1} & 0 & -1 & d_{2} \\
d_{3}d_{0}& 0 & 0 & d_{3}
\end{array} \right | = - I \; ,
\nonumber
\end{eqnarray}

\noindent so that
\begin{eqnarray}
d_{0}=0 \; , \; d_{1}= 0 \; , \; d_{2}=0 \; , \; d_{3}=-1 \;,
\qquad \alpha^{3}= \left | \begin{array}{rrrr}
0 & 0 & 0 & 1 \\
0 & 0 & -1 & 0 \\
0 & 1 & 0 & 0 \\
-1 & 0 & 0 & 0
\end{array} \right | \; , \qquad (\alpha^{3})^{2} = -I\; .
\label{1.8}
\end{eqnarray}

Simple rules for their products hold:
\begin{eqnarray}
\alpha^{1} \alpha^{2} = +\alpha^{3} \; , \; \alpha^{2} \alpha^{1}
= -\alpha^{3} \; , \qquad \alpha^{2} \alpha^{3} = \alpha^{1} \; ,
\; \alpha^{3} \alpha^{2} = -\alpha^{1} \; , \qquad \alpha^{3}
\alpha^{1} = + \alpha^{2}\;, \; \alpha^{1} \alpha^{3} = -
\alpha^{2}\;; \label{1.9}
\end{eqnarray}

\noindent Also we get
\begin{eqnarray}
k=\pm 1 , \qquad  \alpha^{0}\alpha^{1}= \left |
\begin{array}{rrrr}
0 & k & 0 & 0 \\
-1 & 0 & 0 & 0 \\
0 & 0 & 0 & -1 \\
0 & 0 & 1 & 0
\end{array} \right | \; ,\qquad
 \alpha^{1}\alpha^{0}=  \left | \begin{array}{rrrr}
0 & 1 & 0 & 0 \\
-k & 0 & 0 & 0 \\
0 & 0 & 0 & -1 \\
0 & 0 & 1 & 0
\end{array} \right | \; .
\nonumber
\end{eqnarray}

\noindent only when $k=+1$ we get the very simple commutation rule
\begin{eqnarray}
\alpha^{0}= I\; , \qquad \alpha^{i}\alpha^{0} =
\alpha^{0}\alpha^{i}= \alpha^{i} \; . \label{1.9d}
\end{eqnarray}

Thus, Maxwell equations can be presented in the following simple
matrix form
\begin{eqnarray}
(-i \partial_{0} + \alpha^{j} \partial_{j} ) \Psi =J \; , \qquad
\Psi = \left | \begin{array}{c} 0 \\\psi^{1} \\\psi^{2} \\
\psi^{3}
\end{array} \right | \; , \qquad J=
{1 \over \epsilon_{0}} \; \left | \begin{array}{c} j^{0} \\ i\;
j^{1} \\ i\; j^{2} \\ i \; j^{3}
\end{array} \right | \; ,
\nonumber
\\
\alpha^{1} = \left | \begin{array}{rrrr}
0 & 1 & 0 & 0 \\
-1 & 0 & 0 & 0 \\
0 & 0 & 0 & -1 \\
0 & 0 & 1 & 0
\end{array} \right | \; , \qquad
\alpha^{2} = \left | \begin{array}{rrrr}
0 & 0 & 1 & 0 \\
0 & 0 & 0 & 1 \\
-1 & 0 & 0 & 0 \\
0 & -1 & 0 & 0
\end{array} \right | \; , \qquad
\alpha^{3} = \left | \begin{array}{rrrr}
0 & 0 & 0 & 1 \\
0 & 0 & -1 & 0 \\
0 & 1 & 0 & 0 \\
-1 & 0 & 0 & 0
\end{array} \right |\; ,
\nonumber
\\
(\alpha^{1})^{2} = -I \; , \qquad (\alpha^{1})^{2} = -I \; ,
\qquad (\alpha^{1})^{2} = -I \; , \nonumber
\\
\alpha^{1} \alpha^{2}= - \alpha^{2} \alpha^{1} = \alpha^{3} \;,
\qquad \alpha^{2} \alpha^{3} = - \alpha^{3} \alpha^{2} =
\alpha^{1}\;, \qquad \alpha^{3} \alpha^{1} = - \alpha^{1}
\alpha^{3} = \alpha^{2}\;. \label{1.10}
\end{eqnarray}

\section{ Matrix Maxwell equation in Riemannian space
}

\hspace{5mm} Matrix Maxwell equation can be extended to the case
of arbitrary Riemannian space -- time in accordance with the
tetrad approach of Tetrode -- Weyl -- Fock -- Ivanenko:
\begin{eqnarray}
 \alpha^{\rho}(x)\;[ \; \partial_{\rho } + A_{\rho}(x)\; ] \; \Psi (x) = J(x) \; ,
\nonumber
\\
\alpha^{\rho}(x) = \alpha^{c} \; e_{(c)}^{\rho}(x) \; , \qquad
A_{\rho}(x) = {1 \over 2} j^{ab} \; e_{(a)} ^{\beta} \;
\nabla_{\rho} e_{(n) \beta} \; . \label{2.2}
\end{eqnarray}

\noindent where $e_{(c)}^{\rho}(x)$ stands for the tetrad,
$j^{ab}$ stands for -- generators of the complex vector
representation of complex orthogonal group $SO(3.C)$. Eq.
(\ref{2.1}) can be rewritten in terms of rotational Ricci
coefficients
\begin{eqnarray}
\alpha^{c} \; ( \; e_{(c)}^{\rho} \partial_{\rho} + {1 \over 2}
j^{ab} \gamma_{abc} \; ) \; \Psi = J(x) \; , \label{2.3}
\end{eqnarray}

\noindent where $ \gamma_{bac} = - \gamma_{abc} = - e_{(b)\beta
;\alpha} e^{\beta}_{(a)} e^{\alpha}_{(c)} \; $ and
\begin{eqnarray}
j^{23} = s_{1}\; , \qquad j^{01} = i \; s_{1} \; , \qquad j^{31} =
s_{2}\; , \qquad j^{02} = i \; s_{2} \; , \qquad j^{12} = s_{3}\;
, \qquad j^{03} = i \; s_{3} \; , \nonumber
\\
s_{1}= \left | \begin{array}{rrrr}
0 & 0 & 0 & 0 \\
0 & 0 & 0 & 0 \\
0 & 0 & 0 & -1 \\
0 & 0 & 1 & 0 \\
\end{array} \right | \; , \qquad
s_{2} = \left | \begin{array}{rrrr}
0 & 0 & 0 & 0 \\
0 & 0 & 0 & 1 \\
0 & 0 & 0 & 0 \\
0 & -1 & 0 & 0 \\
\end{array} \right | \;,\qquad
s_{3} = \left | \begin{array}{rrrr}
0 & 0 & 0 & 0 \\
0 & 0 & -1 & 0 \\
0 & 1 & 0 & 0 \\
0 & 0 & 0 & 0
\end{array} \right |=
\left | \begin{array}{cc}
0 & 0 \\
0 & \tau_{3}
\end{array} \right | \; \; .
\nonumber
\end{eqnarray}

\section{ Spherical coordinates and tetrad in
the Riemann space $S_{3}$ }

\hspace{5mm} In spherical coordinates in the Riemann space $S_{3}$
(see \cite{Olevskiy})
\begin{eqnarray}
dS^{2}= c^{2} dt^{2}-d \chi^{2}- \sin^{2} \chi \;
(d\theta^{2}+\sin^{2}{\theta}d\phi^{2}) \; , \nonumber
\\
x^{\alpha}=(c t, \chi , \theta, \phi) \; , \qquad
g_{\alpha\beta}=\left |
\begin{array}{cccc}
                  1 & 0 & 0 & 0 \\
                  0 & -1 & 0 & 0 \\
                  0 & 0 & -\sin^{2}\chi & 0 \\
                  0 & 0 & 0 & -\sin^{2}\chi \sin^{2}{\theta}
                        \end {array}
                \right | .
\label{3.1}
\end{eqnarray}

\noindent let us use the following tetrad
\begin{eqnarray}
e^{\alpha}_{(0)}=(1, 0, 0, 0) \; , \qquad e^{\alpha}_{(3)}=(0, 1,
0, 0) \; , \nonumber
\\
e^{\alpha}_{(1)}=(0, 0, \frac {1}{\sin \chi}, 0) \; , \qquad
e^{\alpha}_{(2)}=(1, 0, 0, \frac{1}{ \sin \chi \sin \theta}) \; .
\label{3.2}
\end{eqnarray}

\noindent Christoffel symbols are given by
\begin{eqnarray}
\Gamma^{\chi}_{\phi \phi} = - \sin \chi \cos \chi \sin^{2} \theta
\; , \qquad \Gamma^{\chi}_{\theta \theta} = - \sin \chi \cos \chi
\; , \nonumber
\\
\Gamma^{\theta}_{\phi \phi} = - \sin \theta \cos \theta \; ,
\qquad \Gamma^{\theta}_{\theta \chi} = {\cos \chi \over \sin \chi}
\; , \qquad \Gamma^{\phi}_{\phi \theta} = \mbox{ctg}\; \theta \; ,
\qquad \Gamma^{\phi}_{\chi \phi} = {\cos \chi \over \sin \chi} \;
. \label{3.3}
\end{eqnarray}

\noindent The Ricci coefficients are $\gamma_{ab0} = 0 \; , \;
\gamma_{ab3} = 0\;$ and
\begin{eqnarray}
 \gamma_{ab1}
= \left | \begin{array}{cccc}
0 & 0 & 0 & 0 \\
0 & 0 & 0 & - {1 \over \mbox{tg}\; \chi } \\
0 & 0 & 0 & 0 \\
0 & + {1 \over \mbox{tg} \; \chi } & 0 & 0
\end{array} \right | \; , \qquad
\gamma_{ab2} = \left | \begin{array}{cccc}
0 & 0 & 0 & 0 \\
0 & 0 & + { 1 \over \mbox{tg}\; \theta \; \sin \chi } & 0 \\
0 & - \; { 1 \over \mbox{tg}\; \theta \; \sin \chi } & 0 & - {1 \over \mbox{tg}\; \chi } \\
0 & 0 & + {1 \over \mbox{tg}\; \chi } & 0
\end{array} \right | \; .
\label{3.4}
\end{eqnarray}

\noindent Correspondingly, for $\alpha^{\alpha}(x)$ and
$A_{\alpha}(x)$ we get
\begin{eqnarray}
\alpha^{\alpha}(x) = ( \; \alpha^{0}, \; \alpha^{3}, \; {
\alpha^{1} \over \sin \chi } \; , \; { \alpha^{2} \over \sin \chi
\sin \theta } \; ) \; , \qquad A_{0}(x)=0 \; , \nonumber
\\
  A_{\chi}(x) =0 \; , \qquad A_{\theta}(x) =j^{31} \; , \qquad
A_{\phi} (x) =\sin {\theta} \; j^{32} + \cos {\theta} \; j^{12} \;
. \label{3.5}
\end{eqnarray}

\noindent Therefore, eq. (\ref{2.3}) takes the form
\begin{eqnarray}
[\; -i \partial_{0} \; + \; \alpha^{3}\partial_{r} + {\alpha^{1}
j^{31} + \alpha^{2} j^{32} \over \mbox{tg} \; \chi } \; + \; {1
\over \sin \chi }\; \Sigma_{\theta,\phi } \; ] \; \Psi (x) = 0 \;
, \nonumber
\\
\Sigma_{\theta,\phi} = \alpha^{1} \; \partial_{\theta} \; + \;
\alpha^{2} \; \frac{\partial_{\phi} +
\cos{\theta}j^{12}}{\sin{\theta}} \; . \label{3.6}
\end{eqnarray}

It is more convenient to have the matrix $j^{12}$ as diagonal one.
To this end, one needs to use a cyclic basis
\begin{eqnarray}
 \Psi ' = U_{4} \Psi \; , \qquad
\; U_{4} = \left | \begin{array}{cc} 1 & 0 \\
0 & U
\end{array} \right | \; ,
\label{4.7a}
\end{eqnarray}

\noindent where
\begin{eqnarray}
U = \left |
 \begin{array}{ccc}
- 1 /\sqrt{2} & i / \sqrt{2} & 0 \\[2mm]
0 & 0 & 1 \\[2mm]
1 / \sqrt{2} & i / \sqrt{2} & 0
\end{array} \right | \; , \qquad
U^{-1} = U^{+}_{3} = \left |
 \begin{array}{ccc}
- 1 /\sqrt{2} & 0 & 1 / \sqrt{2} \\[2mm]
-i / \sqrt{2} & 0 & -i / \sqrt{2} \\[2mm]
0 & 1 & 0
\end{array} \right | \; .
\nonumber
\end{eqnarray}

\noindent In is matter of simple calculation to find
\begin{eqnarray}
U \tau_{1} U^{-1} = {1 \over \sqrt{2}} \left |
\begin{array}{ccc}
0 & -i & 0 \\
-i & 0 & -i \\
0 & -i & 0
\end{array} \right | = \tau'_{1} \; ,
\qquad j^{'23} = s_{1}' = \left | \begin{array}{cc}
0 & 0 \\
0 & \tau'_{1}
\end{array} \right | \; ,
\nonumber
\\
U \tau_{2} U^{-1} = {1 \over \sqrt{2}} \left |
\begin{array}{ccc}
0 & -1 & 0 \\
1 & 0 & -1 \\
0 & 1 & 0
\end{array} \right | = \tau'_{2}\; , \qquad j^{'31} = s_{2}' = \left | \begin{array}{cc}
0 & 0 \\
0 & \tau'_{2}
\end{array} \right | \;
, \nonumber
\\
U \tau_{3} U^{-1} = - i \; \left | \begin{array}{rrr}
+1 & 0 & 0 \\
0 & 0 & 0 \\
0 & 0 & -1
\end{array} \right | = \tau'_{3} \; \qquad j^{'12} = s_{3}' = \left | \begin{array}{cc}
0 & 0 \\
0 & \tau'_{3}
\end{array} \right | \;
. \; \nonumber
\end{eqnarray}
\begin{eqnarray}
\ \alpha^{'1} = {1 \over \sqrt{2}} \left |
\begin{array}{rrrr}
0 & - 1 & 0 & 1 \\
1 & 0 & -i & 0 \\
0 & -i & 0 & - i \\
-1 & 0 & -i & 0
\end{array} \right | \; ,\;
\alpha^{'2} = {1 \over \sqrt{2}} \left | \begin{array}{rrrr}
0 & - i & 0 & -i \\
-i & 0 & -1 & 0 \\
0 & 1 & 0 & -1 \\
-i & 0 & 1 & 0
\end{array} \right | \; ,\;
 \alpha^{'3} \; = \; \left |
\begin{array}{rrrr}
0 & 0 & 1 & 0 \\
0 & -i & 0 & 0 \\
-1 & 0 & 0 & 0 \\
0 & 0 & 0 & +i
\end{array} \right | \; .
\nonumber
\end{eqnarray}

\noindent In cyclic basis, eq. (\ref{3.6}) reads
\begin{eqnarray}
[\; -i {\partial \over \partial t } \; + \; \alpha^{'3} {\partial
\over \partial r} + {\alpha^{'1} s'_{2} - \alpha^{'2} s'_{1} \over
\mbox{tg} \; \chi } \; + \; {1 \over \sin \chi }\;
\Sigma'_{\theta,\phi } \; ] \; \Psi '(x) = 0 \; , \nonumber
\\
\Sigma'_{\theta,\phi} = \alpha^{'1} \; \partial_{\theta} \; + \;
\alpha^{'2} \; \frac{\partial_{\phi} + \cos{\theta} \; s'_{3}
}{\sin{\theta}} \; . \label{3.8}
\end{eqnarray}

\section{ Separation of variables and Wigner functions}

\hspace{5mm} Let us construct electromagnetic spherical waves,
then the field function should be taken in the form
\begin{eqnarray}
\psi = e^{-i \omega t} \left | \begin{array}{l}
0 \\
f_{1}(r) D_{-1 }
\\
f_{2}(r) D_{0 } \\
f_{3}(r) D_{+1 }
\end{array} \right |
\label{4.1}
\end{eqnarray}

\noindent where the Wigner $D$-function are used $ D_{\sigma} =
D^{j}_{-m, \sigma} ( \phi , \theta, 0)\;, \;\; \sigma = -1, 0, +1$
; $j,m$ determine ${\bf J}^{2}$ and $J_{3}$ eigenvalues. When
separating the variables we will need the recurrent relations for
Wigner's function \cite{Varsh}:
\begin{eqnarray}
\partial_{\theta} \; D_{-1} = {1 \over 2} \; ( \;
a \; D_{-2} - \nu \; D_{0} \; ) \; , \qquad \frac {m -
\cos{\theta}}{\sin {\theta}} \; D_{-1} = {1 \over 2} \; ( \; a \;
D_{-2} + \nu \; D_{0} \; ) \; , \nonumber
\\
\partial_{\theta} \; D_{0} = {1 \over 2} \; ( \;
\nu \; D_{-1} - \nu \; D_{+1} \; ) \; , \qquad \frac {m}{\sin
{\theta}} \; D_{0} = {1 \over 2} \; ( \; \nu \; D_{-1} + \nu \;
D_{+1} \; ) \; , \nonumber
\\
\partial_{\theta} \; D_{+1} =
{1 \over 2} \; ( \; \nu \; D_{0} - a \; D_{+2} \; ) \; , \qquad
\frac {m + \cos{\theta}}{\sin {\theta}} \; D_{+1} ={1 \over 2} \;
( \;
 \nu \; D_{0} + \ a \; D_{+2} \; ) \; ,
\nonumber
\\
\nu = \sqrt{j(j+1)} \; , \;\; a = \sqrt{(j-1)(j+2)} \; .
\label{4.2}
\end{eqnarray}

Let us find the action of the angular operator (the factor $e^{-i
\omega t} $ will be omitted for shortness)
\begin{eqnarray}
\sqrt{2}\; \Sigma'_{\theta \phi} \; \Psi '= \sqrt{2} \; \left [ \;
\alpha^{'1} \; \partial_{\theta} \; + \; \alpha^{'2} \;
\frac{\partial_{\phi} + \cos{\theta} \; s'_{3} }{\sin{\theta}} \;
\right ] \left | \begin{array}{c}
0 \\
f_{1}(r) D_{-1 }
\\
f_{2}(r) D_{0 } \\
f_{3}(r) D_{+1 }
\end{array} \right | =
\nonumber
\\
\left | \begin{array}{rrrr}
0 & - 1 & 0 & 1 \\
1 & 0 & -i & 0 \\
0 & -i & 0 & - i \\
-1 & 0 & -i & 0
\end{array} \right | \partial_{\theta}
\; \left | \begin{array}{c}
0 \\
f_{1} D_{-1 }
\\
f_{2} D_{0 } \\
f_{3} D_{+1 }
\end{array} \right |+
  {1 \over \sin \theta } \; \left | \begin{array}{rrrr}
0 & - i & 0 & -i \\
-i & 0 & -1 & 0 \\
0 & 1 & 0 & -1 \\
-i & 0 & 1 & 0
\end{array} \right | \;
\left | \begin{array}{c}
0 \\
f_{1} ( im - i \cos \theta )\; D_{-1 }
\\
f_{2} ( i m ) \; D_{0 } \\
f_{3} ( im + i \cos \theta ) \; D_{+1 }
\end{array} \right | \; =
\nonumber
\\
= {1 \over 2} \left | \begin{array}{rrrr}
0 & - 1 & 0 & 1 \\
1 & 0 & -i & 0 \\
0 & -i & 0 & - i \\
-1 & 0 & -i & 0
\end{array} \right |
\; \left | \begin{array}{c}
0 \\
f_{1} (  a  D_{-2} - \nu  D_{0}  )
\\
f_{2} (
\nu  D_{-1} - \nu  D_{+1}  ) \\
f_{3} (  \nu  D_{0} - a  D_{+2}  )
\end{array} \right |
+
 {1 \over 2}
  \left | \begin{array}{rrrr}
0 & +1 & 0 & +1 \\
+1 & 0 & -i & 0 \\
0 & i & 0 & -i \\
+1 & 0 & i & 0
\end{array} \right | \;
\left | \begin{array}{c}
0 \\
f_{1} (  a  D_{-2} + \nu  D_{0}  )
\\
f_{2} (  \nu
D_{-1} + \nu  D_{+1}  ) \\
f_{3} ( \;
 \nu  D_{0} +  a  D_{+2}  )
\end{array} \right |
\nonumber
\end{eqnarray}

\noindent from whence we arrive at
\begin{eqnarray}
 \Sigma'_{\theta \phi} \Psi ' =
 { \nu \over \sqrt {2}} \;
 \left | \begin{array}{r}
 (f_{1} +f_{3}) D_{0} \\
 -i \; f_{2} D_{-1}
 \\
 i \; (f_{1} -f_{3}) D_{0}
 \\
 + i \; f_{2} D_{+1}
 \end{array} \right |
 \label{4.3}
\end{eqnarray}

From Maxwell equation,
 taking into account (\ref{4.3}) and identities
\begin{eqnarray}
 -i \partial_{0} \Psi =
 - \omega \; e^{-i\omega t} \; \left | \begin{array}{c}
0 \\
f_{1}(r) D_{-1 }
\\
f_{2}(r) D_{0 } \\
f_{3}(r) D_{+1 }
\end{array} \right | \; , \qquad
\alpha^{'3}\partial_{r} \Psi ' = e^{-i\omega t} \;\; \left |
\begin{array}{c}
f'_{2} \; D_{0} \\
-i \; f'_{1} \; D_{-1} \\
0 \\
+ i \; f'_{3} \; D_{+1}
\end{array} \right | \; ,
\nonumber
\\
{\alpha^{'1} s'_{2} - \alpha^{'2} s'_{1} \over \mbox{tg} \; \chi }
\; \Psi ' = {e^{-i\omega t} \over \mbox{tg}\; \chi } \; \; \left |
\begin{array}{cccc}
0 & 0 & 2 & 0 \\
0 & -i & 0 & 0 \\
0 & 0 & 0 & 0 \\
0 & 0 & 0 & +i
\end{array} \right |
\left | \begin{array}{c}
0 \\
f_{1}(r) D_{-1 }
\\
f_{2}(r) D_{0 } \\
f_{3}(r) D_{+1 }
\end{array} \right | =
{e^{-i\omega t} \over \mbox{tg}\; \chi } \; \left |
\begin{array}{c}
2f_{2}(r) D_{0 } \\
-i \; f_{1}(r) D_{-1 }
\\
0 \\
+i \; f_{3}(r) D_{+1 }
\end{array} \right |,
\label{4.4}
\end{eqnarray}

\noindent we get the radial equations
\begin{eqnarray}
f'_{2} + {2 \over \mbox{tg}\; \chi} f_{2} + {1 \over \sin \chi }
\;
 { \nu \over \sqrt {2}} \; (f_{1} +f_{3})= 0\; ,
\nonumber
\\
- \omega f_{1} - i\; f_{1} ' - { i \over \mbox{tg}\; \chi } f_{1}
-
 {i \over \sin \chi } \; { \nu \over \sqrt {2}} \; f_{2} = 0 \; ,
\nonumber
\\
- \omega f_{2} + {i \over \sin \chi } \;
 { \nu \over \sqrt {2}} (f_{1} -f_{3}) = 0 \; ,
\nonumber
\\
- \omega f_{3} + i\; f_{3}' + {i \over \mbox{tg}\; \chi} f_{3} +
{i \over \sin \chi } \; { \nu \over \sqrt {2}} \; f_{2} = 0 \; .
\label{4.5}
\end{eqnarray}

With the use of substitutions
\begin{eqnarray}
f_{1} = {1 \over \sin \chi }\; F_{1} \; \qquad
 f_{2} = {1 \over \sin \chi }\; F_{2} \; , \qquad
 f_{3} = {1 \over \sin \chi }\; F_{3} \; ,
\nonumber
\end{eqnarray}

\noindent the systems reads simpler
\begin{eqnarray}
(1) \qquad ( {d \over d \chi } + {1 \over \mbox{tg}\; \chi}) \;
\omega F_{2} +
 { \omega\; \nu \over \sqrt {2} \sin \chi } \; (F_{1} +F_{3})= 0\; ,
\nonumber
\\
(2) \qquad - \omega^{2} F_{1} - i \omega \; F_{1} ' -
   { i \; \nu \over \sqrt {2} \sin \chi } \; \omega F_{2} = 0 \; ,
\nonumber
\\
(3) \qquad \omega F_{2} =
 { i \nu \over \sqrt {2} \sin \chi } (F_{1} -F_{3}) \; ,
\nonumber
\\
(4 ) \qquad - \omega^{2} F_{3} + i\; \omega F_{3}' + { i \nu \over
\sqrt {2} \sin \chi } \; \omega F_{2} = 0 \; . \label{5.6}
\end{eqnarray}

Combining eqs. (2) and (4) we get

\vspace{1mm} $ \;\;\;(2) + (4)\; , $
\begin{eqnarray}
- \omega (F_{1} + F_{3}) - i ( F'_{1}- F'_{3}) = 0 \label{5.7a}
\end{eqnarray}

\vspace{1mm} $ -(2) + (4)\; , $
\begin{eqnarray}
\omega^{2} (F_{1} - F_{3} ) + i \omega (F_{1}' + F_{3}') + { 2 i
\nu \over \sqrt{2} \sin \chi }\;
 { i \nu \over \sqrt {2} \sin \chi } (F_{1} -F_{3})= 0
 \label{4.7b}
\end{eqnarray}

Allowing for eq. (3) and (\ref{5.7a}), from eq. (3) in (\ref{5.6})
we arrive at an identity
\begin{eqnarray}
( {d \over d \chi } + {1 \over \mbox{tg}\; \chi}) \; { i \nu \over
\sqrt {2} \sin \chi } (F_{1} -F_{3}) +
 { \omega\; \nu \over \sqrt {2} \sin \chi } \; (-i) (F'_{1} -F'_{3})= 0\; .
\label{4.8}
\end{eqnarray}

Therefore, the Maxwell equations reduce to only three independent
equations:
\begin{eqnarray}
\omega F_{2} =
 { i \nu \over \sqrt {2} \sin \chi } (F_{1} -F_{3}) \; , \qquad
 - \omega (F_{1} + F_{3}) - i ( F'_{1}- F'_{3}) = 0 \; ,
\nonumber
\\
\omega^{2} (F_{1} - F_{3} ) + i \omega (F_{1}' + F_{3}') + { 2 i
\nu \over \sqrt{2} \sin \chi }\;
 { i \nu \over \sqrt {2} \sin \chi } (F_{1} -F_{3})= 0 \; .
\label{4.9}
\end{eqnarray}

\noindent Let us introduce new variables:
\begin{eqnarray}
F = {F_{1} + F_{3} \over \sqrt{2} } \; , \qquad G = {F_{1}- F_{3}
\over \sqrt{2} } \; , \nonumber
\end{eqnarray}

\noindent then (\ref{4.9}) will read
\begin{eqnarray}
F_{2} =
 { i \nu \over \omega \; \sin \chi } \; G \; , \qquad
  F = - { i \over \omega }\; {d \over d \chi } G \; , \qquad
  {d ^{2} \over d \chi^{2}} \; G + \omega^{2} G - { \nu^{2} \over
\sin^{2} \chi }\; G = 0 \; . \label{4.10}
\end{eqnarray}

\section { Solution of the radial equation in $S_{3}$ }

\hspace{5mm} Let us solve eq. (\ref{4.10}). To this end one need
to introduce a new variable
\begin{eqnarray}
z = 1 - e^{-2i\chi} \; , \qquad z = 2 \sin \chi \; e^{i(-\chi +
\pi /2)} \; ;  \label{5.1}
\end{eqnarray}

\noindent $z$ runs along closed path in the complex plane:

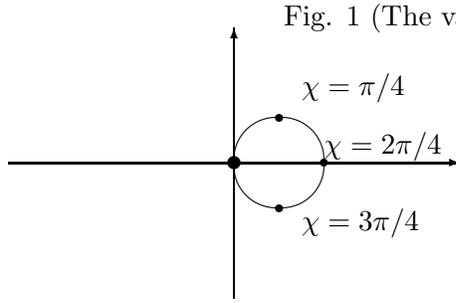
\begin{center}
Fig. 1 (The variable $z$)
\end{center}

\vspace{-10mm} \unitlength=0.6mm
\begin{picture}(160,40)(-80,0)
\special{em:linewidth 0.4pt} \linethickness{0.4pt}

\put(-50,0){\vector(+1,0){100}} \put(0,-30){\vector(0,+1){60}}
\put(+10,0){\oval(20,20)}

\put(0,0){\circle*{3}} \put(+20,0){\circle*{2}}
\put(+10,+10){\circle*{2}} \put(+15,+15){$\chi = \pi /4$}
\put(+20,+2){$\chi = 2\pi /4$} \put(+10,-10){\circle*{2}}
\put(+15,-15){$\chi = 3 \pi /4$}

\end{picture}

\vspace{20mm}

Allowing for identities
\begin{eqnarray}
{d \over d \chi} = 2i (1 - z) {d \over d z} \; , \qquad {\cos \chi
\over \sin \chi } = i {2 - z \over z } \; , \qquad {1 \over
\sin^{2} \chi } = -{4(1-z) \over z^{2} } \; , \nonumber
\end{eqnarray}

\noindent eq. (\ref{4.10}) reduces to
\begin{eqnarray}
4(1-z)^{2}{d ^{2} G\over dz^{2}} \; -4 (1-z){dG\over dz}-
\omega^{2} G -{4(1-z)\nu^{2} \over z^{2} }\; G = 0 \; .
\label{5.2}
\end{eqnarray}

\noindent With the use of the substitution
\begin{eqnarray}
G = z^{a} (1-z)^{b} g (z)\;, \nonumber
\\
G'=a z^{a-1} (1-z)^{b} g (z)-bz^{a} (1-z)^{b-1} g (z)+z^{a}
(1-z)^{b} {dg (z)\over dz}\;, \nonumber
\\
G''=a (a-1)z^{a-2} (1-z)^{b}g (z)-abz^{a-1} (1-z)^{b-1} g
(z)+az^{a-1} (1-z)^{b} {dg (z)\over dz}- \nonumber
\\
-a bz^{a-1} (1-z)^{b-1} g (z)+b(b-1)z^{a} (1-z)^{b-2} g (z)-bz^{a}
(1-z)^{b-1} {dg (z)\over dz}+ \nonumber
\\
+a z^{a-1} (1-z)^{b} {dg (z)\over dz}-bz^{a} (1-z)^{b-1} {dg
(z)\over dz}+z^{a} (1-z)^{b} {d^{2}g (z)\over
dz^{2}}\;.\label{5.3}
\end{eqnarray}

\noindent from (\ref{5.2}) we arrive at
\begin{eqnarray}
z (1-z) {d^{2}g \over dz^{2}} + [ 2a -(2a+2b+1)z]\; {d g\over d z}
+ \nonumber
\\
 + \left [ {\omega^{2} \over 4}-(a+b)^{2} +(a(a-1)-\nu^{2}){1 \over
z} + (b^{2}-{\omega^{2} \over 4}){1 \over 1-z} \right ] g=0.
\nonumber
\end{eqnarray}

\noindent Requiring
\begin{eqnarray}
a(a-1)-\nu^{2} = 0 \; , \qquad b^{2}-{\omega^{2} \over 4}=0
\qquad  \Longrightarrow \qquad  a = j+1 , - j \; , \qquad b = \pm
{\omega \over 2} \; , \label{6.3}
\end{eqnarray}

\noindent for $g$ we obtain a simpler equation
\begin{eqnarray}
z (1-z) {d^{2}g \over dz^{2}} + \left [ 2a -(2a+2b+1)z\right]\; {d
g\over d z}
 - \left [(a+b)^{2} - {\omega^{2} \over 4} \right ] g=0 \; ,
\label{6.4}
\end{eqnarray}

\noindent which is of hypergeometric type
\begin{eqnarray}
z(1-z) \; F'' + [ \gamma - (\alpha + \beta +1) z ] \; F' - \alpha
\beta \; F = 0 \nonumber
\end{eqnarray}

\noindent with parameters
\begin{eqnarray}
\gamma = 2a \; , \qquad \alpha + \beta = 2a + 2b \;, \qquad \alpha
\beta = (a+b)^{2} - {\omega^{2} \over 4} \; , \nonumber
\end{eqnarray}

\noindent or
\begin{eqnarray}
\alpha = a+b - { \omega \over 2} \;, \qquad \beta = a+b + {\omega
\over 2} \; . \label{5.5}
\end{eqnarray}

\noindent The function $G$ is given by
\begin{eqnarray}
G = z^{a} (1- z)^{b} g(z) = \left [\; 2i \sin \chi e^{-i \chi } \;
\right ] ^{a} \; \left [ \; 1 - 2i \sin \chi e^{-i \chi } \;
\right ] ^{b} \; g(z)\; ; \label{5.6}
\end{eqnarray}

\noindent it is finite at the points $\chi =0$ and $\chi = \pi$
only when $a$ is positive (see (\ref{6.3})):
\begin{eqnarray}
a = j+1 \; . \label{5.7}
\end{eqnarray}

\noindent also we must require $b = - \omega /2 $ when
hypergeometric series can be reduced to a polynomial
 (we take $\omega > 0$):
\begin{eqnarray}
\alpha = j+1 - \omega = - n = \{ 0, -1, -2, ... \; \} \qquad
\Longrightarrow \qquad \omega = n + 1 + j \; ; \label{5.8}
\end{eqnarray}

Thus, physical solutions of the Maxwell equations in the Riemann
space $S_{3}$ are given by relations:
\begin{eqnarray}
G = z^{a} (1- z)^{b} g(z) = \left [\; 2i \sin \chi e^{-i \chi } \;
\right ] ^{a} \; \left [ \; 1 - 2i \sin \chi e^{-i \chi } \;
\right ] ^{b} \; g(z)\; , \nonumber
\\
g(z) = F (-n, , j +1, 2j +2; \; z) = F (-n, , j +1, 2j +2; 2i \sin
\chi e^{-i \chi } ) \label{5.9a}
\end{eqnarray}

\noindent where
\begin{eqnarray}
\omega = n + 1 + j \; , \qquad j = 0, 1, 2, ...., \qquad n = 0, 1,
2, ... ; \label{5.9b}
\end{eqnarray}

\noindent or in usual units
\begin{eqnarray}
\omega = {c \over \rho } \; ( n + 1 + j )\; , \label{5.9c}
\end{eqnarray}

\noindent $\rho$ stands for the curvature radius of the space, $c$
is velocity of the light.

\section{ Spherical coordinates and tetrad in Lobachevsky space $H_{3}$
}

\hspace{5mm} Let us consider Maxwell equation in spherical
coordinates of the Lobachevsky space model $H_{3}$
\begin{eqnarray}
dS^{2}= c^{2} dt^{2}-d \chi^{2}- \mbox{sh}^{2} \chi \;
(d\theta^{2}+\sin^{2}{\theta}d\phi^{2}) \; , \nonumber
\\
x^{\alpha}=(c t, \chi , \theta, \phi) \; , \qquad
g_{\alpha\beta}=\left |
\begin{array}{cccc}
                  1 & 0 & 0 & 0 \\
                  0 & -1 & 0 & 0 \\
                  0 & 0 & -\mbox{sh}^{2}\chi & 0 \\
                  0 & 0 & 0 & -\mbox{sh}^{2}\chi \sin^{2}{\theta}
                        \end {array}
                \right | .
 \label{6.1}
\end{eqnarray}

\noindent in the following tetrad
\begin{eqnarray}
e^{\alpha}_{(0)}=(1, 0, 0, 0) \; , \qquad e^{\alpha}_{(3)}=(0, 1,
0, 0) \; , \nonumber
\\
e^{\alpha}_{(1)}=(0, 0, \frac {1}{\mbox{sh}\; \chi}, 0) \; ,
\qquad e^{\alpha}_{(2)}=(1, 0, 0, \frac{1}{ \mbox{sh}\; \chi \sin
\theta}) \; . \label{6.2}
\end{eqnarray}

\noindent The Christoffel symbols are
\begin{eqnarray}
\Gamma^{\chi}_{\phi \phi} = - \mbox{sh}\; \chi \mbox{ch}\; \chi
\sin^{2} \theta \; , \qquad \Gamma^{\chi}_{\theta \theta} = -
\mbox{sh}\; \chi \mbox{ch}\; \chi \; , \nonumber
\\
\Gamma^{\theta}_{\phi \phi} = - \sin \theta \cos \theta \; ,
\qquad \Gamma^{\theta}_{\theta \chi} = {\mbox{ch}\; \chi \over
\mbox{sh}\; \chi} \; , \qquad \Gamma^{\phi}_{\phi \theta} =
\mbox{ctg}\; \theta \; , \qquad \Gamma^{\phi}_{\chi \phi} =
{\mbox{ch}\; \chi \over \mbox{sh}\; \chi} \; . \label{6.3a}
\end{eqnarray}

\noindent and the Ricci rotation coefficients are $\gamma_{ab0} =
0 \; , \; \gamma_{ab3} = 0$ and
\begin{eqnarray}
\gamma_{ab1} = \left | \begin{array}{cccc}
0 & 0 & 0 & 0 \\
0 & 0 & 0 & - {1 \over \mbox{th}\; \chi } \\
0 & 0 & 0 & 0 \\
0 & + {1 \over \mbox{th} \; \chi } & 0 & 0
\end{array} \right | \; , \qquad
\gamma_{ab2} = \left | \begin{array}{cccc}
0 & 0 & 0 & 0 \\
0 & 0 & + { 1 \over \mbox{tg}\; \theta \; \mbox{sh}\; \chi } & 0 \\
0 & - \; { 1 \over \mbox{tg}\; \theta \; \mbox{sh}\; \chi } & 0 & - {1 \over \mbox{th}\; \chi } \\
0 & 0 & + {1 \over \mbox{th}\; \chi } & 0
\end{array} \right | \; .
\label{6.4}
\end{eqnarray}

\noindent For $\alpha^{\alpha}(x)$ and $A_{\alpha}(x)$ we get
\begin{eqnarray}
\alpha^{\alpha}(x) = ( \; \alpha^{0}, \; \alpha^{3}, \; {
\alpha^{1} \over \mbox{sh}\; \chi } \; , \; { \alpha^{2} \over
\mbox{sh}\; \chi \sin \theta } \; ) \; , \qquad A_{0}(x)=0 \; ,
\nonumber
\\
  A_{\chi}(x) =0 \; , \qquad A_{\theta}(x) =j^{31} \; , \qquad
A_{\phi} (x) =\sin {\theta} \; j^{32} + \cos {\theta} \; j^{12} \;
. \label{6.5}
\end{eqnarray}

\noindent Therefore, Maxwell equation  reads  (the cyclic will be
used)
\begin{eqnarray}
[\; -i {\partial \over \partial t } \; + \; \alpha^{'3} {\partial
\over \partial r} + {\alpha^{'1} s'_{2} - \alpha^{'2} s'_{1} \over
\mbox{th} \; \chi } \; + \; {1 \over \mbox{sh}\; \chi }\;
\Sigma'_{\theta,\phi } \; ] \; \Psi '(x) = 0 \; , \nonumber
\\
\Sigma'_{\theta,\phi} = \alpha^{'1} \; \partial_{\theta} \; + \;
\alpha^{'2} \; \frac{\partial_{\phi} + \cos{\theta} \; s'_{3}
}{\sin{\theta}} \; . \label{6.7}
\end{eqnarray}

\section{ Separation of variables
}

\hspace{5mm} We start with spherical substitution
\begin{eqnarray}
\psi = e^{-i \omega t} \left | \begin{array}{l}
0 \\
f_{1}(r) D_{-1 }
\\
f_{2}(r) D_{0 } \\
f_{3}(r) D_{+1 }
\end{array} \right |
\label{7.1}
\end{eqnarray}

\noindent Further calculations are completely the same that were
used in previous case, so that we can go just to the final result:
\begin{eqnarray}
f'_{2} + {2 \over \mbox{th}\; \chi} f_{2} + {1 \over \mbox{sh}\;
\chi } \;
 { \nu \over \sqrt {2}} \; (f_{1} +f_{3})= 0\; ,
\nonumber
\\
- \omega f_{1} - i\; f_{1} ' - { i \over \mbox{th}\; \chi } f_{1}
-
 {i \over \mbox{sh}\; \chi } \; { \nu \over \sqrt {2}} \; f_{2} = 0 \; ,
\nonumber
\\
- \omega f_{2} + {i \over \mbox{sh}\; \chi } \;
 { \nu \over \sqrt {2}} (f_{1} -f_{3}) = 0 \; ,
\nonumber
\\
- \omega f_{3} + i\; f_{3}' + {i \over \mbox{th}\; \chi} f_{3} +
{i \over \mbox{sh}\; \chi } \; { \nu \over \sqrt {2}} \; f_{2} = 0
\; . \label{7.2}
\end{eqnarray}

\noindent The system becomes simpler with the substitution
\begin{eqnarray}
f_{1} = {1 \over \mbox{sh}\; \chi }\; F_{1} \;, \qquad
 f_{2} = {1 \over \mbox{sh}\; \chi }\; F_{2} \; , \qquad
 f_{3} = {1 \over \mbox{sh}\; \chi }\; F_{3} \; ,
\nonumber
\end{eqnarray}

\noindent so we tet to
\begin{eqnarray}
(1) \qquad ( {d \over d \chi } + {1 \over \mbox{th}\; \chi}) \;
\omega F_{2} +
 { \omega\; \nu \over \sqrt {2} \mbox{sh}\; \chi } \; (F_{1} +F_{3})= 0\; ,
\nonumber
\\
(2) \qquad - \omega^{2} F_{1} - i \omega \; F_{1} ' -
   { i \; \nu \over \sqrt {2} \mbox{sh}\; \chi } \; \omega F_{2} = 0 \; ,
\nonumber
\\
(3) \qquad \omega F_{2} =
 { i \nu \over \sqrt {2} \mbox{sh}\; \chi } (F_{1} -F_{3}) \; ,
\nonumber
\\
(4 ) \qquad - \omega^{2} F_{3} + i\; \omega F_{3}' + { i \nu \over
\sqrt {2} \mbox{sh}\; \chi } \; \omega F_{2} = 0 \; . \label{7.3}
\end{eqnarray}

\noindent Combining (2) и (4) we get

\vspace{3mm} $ \;\;\; (2) + (4)\; , $
\begin{eqnarray}
- \omega (F_{1} + F_{3}) - i ( F'_{1}- F'_{3}) = 0 \label{7.4a}
\end{eqnarray}

\vspace{3mm} $ -(2) + (4)\; , $
\begin{eqnarray}
\omega^{2} (F_{1} - F_{3} ) + i \omega (F_{1}' + F_{3}') + { 2 i
\nu \over \sqrt{2} \mbox{sh}\; \chi }\;
 { i \nu \over \sqrt {2} \mbox{sh}\; \chi } (F_{1} -F_{3})= 0
\label{7.4b}
\end{eqnarray}

\noindent eq. (1) in (\ref{7.3}) reduce to identity $0 \equiv 0$
when allowing for (3) and и (\ref{7.4a}), то придем к тождеству $0
\equiv 0$. So we have only three independent equations:
\begin{eqnarray}
\omega F_{2} =
 { i \nu \over \sqrt {2} \mbox{sh}\; \chi } (F_{1} -F_{3}) \; , \qquad
- \omega (F_{1} + F_{3}) - i ( F'_{1}- F'_{3}) = 0 \; , \nonumber
\\
\omega^{2} (F_{1} - F_{3} ) + i \omega (F_{1}' + F_{3}') + { 2 i
\nu \over \sqrt{2} \mbox{sh}\; \chi }\;
 { i \nu \over \sqrt {2} \mbox{sh}\; \chi } (F_{1} -F_{3})= 0 \; .
\label{7.5}
\end{eqnarray}

\noindent In new field variables
\begin{eqnarray}
F = {F_{1} + F_{3} \over \sqrt{2} } \; , \qquad G = {F_{1}- F_{3}
\over \sqrt{2} } \; , \nonumber
\end{eqnarray}

\noindent eqs. (\ref{7.5}) read
\begin{eqnarray}
F_{2} =
 { i \nu \over \omega \; \mbox{sh}\; \chi } \; G \; , \qquad
  F = - { i \over \omega }\; {d \over d \chi } G \; , \qquad
  {d ^{2} \over d \chi^{2}} \; G + \omega^{2} G - { \nu^{2} \over
\mbox{sh}\;^{2} \chi }\; G = 0 \; . \label{7.6}
\end{eqnarray}

\section { Solution of the radial equations in $H_{3}$}

\hspace{5mm} In eq. (\ref{7.6}) one needs to introduce new
variable
\begin{eqnarray}
z = 1 - e^{-2\chi} \; , \qquad z = 2 \mbox{sh}\; \chi \; e^{-\chi
} \; . \label{8.1}
\end{eqnarray}

\noindent Allowing for identities
\begin{eqnarray}
{d \over d \chi} = 2 (1 - z) {d \over d z} \; , \qquad
{\mbox{ch}\; \chi \over \mbox{sh}\; \chi } = {2 - z \over z } \; ,
\qquad {1 \over \mbox{sh}\;^{2} \chi } = {4(1-z) \over z^{2} } \;
, \nonumber
\end{eqnarray}

\noindent eq. (\ref{7.6}) reduces to
\begin{eqnarray}
4(1-z)^{2}{d ^{2} G\over dz^{2}} \; -4 (1-z){dG\over dz}+
\omega^{2} G -{4(1-z)\nu^{2} \over z^{2} }\; G = 0 \; .
\label{8.2}
\end{eqnarray}

With the use of the substitution
\begin{eqnarray}
G = z^{a} (1-z)^{b} g (z)\;, \nonumber
\end{eqnarray}

\noindent we arrive at
\begin{eqnarray}
z (1-z) {d^{2}g \over dz^{2}} + [ 2a -(2a+2b+1)z]\; {d g\over d z}
+ \nonumber
\\
 + \left [- {\omega^{2} \over 4}-(a+b)^{2} +(a(a-1)-\nu^{2}){1 \over
z} + (b^{2}+{\omega^{2} \over 4}){1 \over 1-z} \right ] g=0 \; .
\nonumber
\end{eqnarray}

\noindent With additional restrictions
\begin{eqnarray}
a(a-1)-\nu^{2} = 0 \; , \qquad b^{2}+{\omega^{2} \over 4}=0 \qquad
\Longrightarrow \qquad  a = j+1 , - j \; , \qquad b = \pm {i\omega
\over 2} \; . \label{8.3}
\end{eqnarray}

\noindent for $g$ we obtain equation
\begin{eqnarray}
z (1-z) {d^{2}g \over dz^{2}} + \left [ 2a -(2a+2b+1)z\right]\; {d
g\over d z}
 - \left [(a+b)^{2} + {\omega^{2} \over 4} \right ] g=0 ,
\label{8.4}
\end{eqnarray}

\noindent of hypergeometric type with parameters
\begin{eqnarray}
\gamma = 2a \; , \qquad \alpha + \beta = 2a + 2b \;, \qquad \alpha
\beta = (a+b)^{2} + {\omega^{2} \over 4} \; , \nonumber
\end{eqnarray}

\noindent or
\begin{eqnarray}
\alpha = a+b - { i\omega \over 2} \;, \qquad \beta = a+b +
{i\omega \over 2} \; . \label{8.5}
\end{eqnarray}

\noindent Therefore, the function $G$ is given by
\begin{eqnarray}
G = z^{a} (1- z)^{b} g(z) = \left [\; 2 \mbox{sh}\; \chi e^{- \chi
} \; \right ] ^{a} \; \left [ \; 1 - 2 \mbox{sh}\; \chi e^{- \chi
} \; \right ] ^{b} \; g(z)\; . \label{8.6}
\end{eqnarray}

\section{ Conclusions}

Complex formalism of Riemann -- Silberstein -- Majorana --
Oppenheimer in Maxwell electro\- dynamics is extended to the case
of arbitrary pseudo-Riemannian space - time in accordance with the
tetrad recipe of Tetrode -- Weyl -- Fock -- Ivanenko. In this
approach, the Maxwell equations are solved exactly
 on the background of simplest static cosmological models, spaces of constant curvature of Riemann and Lobachevsky
  parameterized by spherical coordinates.
Separation of variables is realized in the basis of
Schr\"{o}dinger -- Pauli type, description of angular dependence
in electromagnetic complex 3-vectors is given in terms of Wigner
$D$-functions. In the case of compact Riemann model a discrete
frequency spectrum
 for electromagnetic modes depending on the curvature radius of space and three discrete
 parameters is found. In the case of hyperbolic Lobachevsky model
 no discrete spectrum for frequencies of electromagnetic modes arises.

Authors are thankful to all participant of scientific seminar of
Laboratory of Theoretical Physics of Institute of Physic on NASB
for discussion and advice.

\end{document}